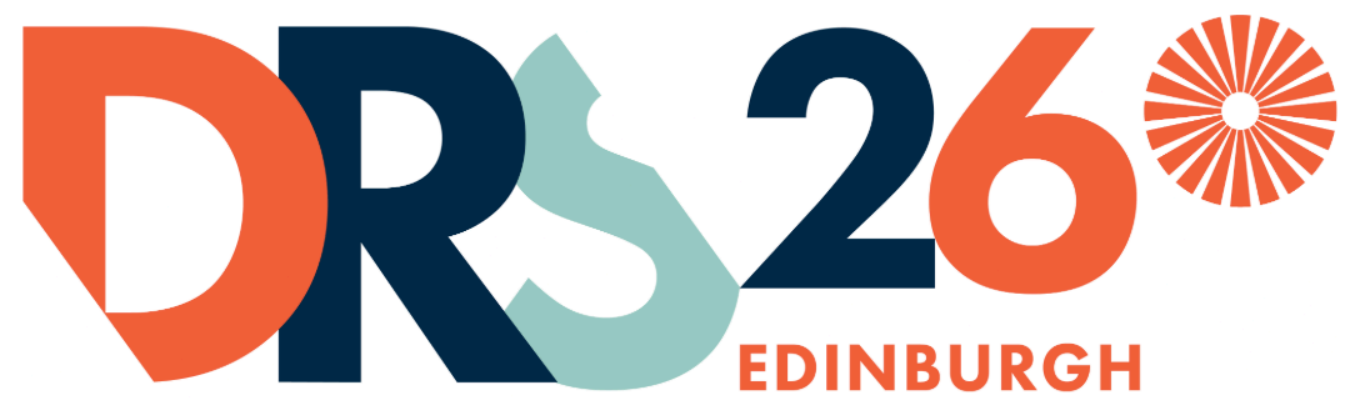

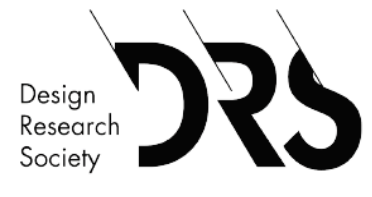

# Framing Data Choices: How Pre-Donation Exploration Designs Influence Data Donation Behavior and Decision-Making

Zeya Chen*, Zach Pino, Ruth Schmidt

Institute of Design (ID) at Illinois Institute of Technology, Chicago, USA

*Corresponding author e-mail: zchen103@hawk.illinoistech.edu

**Abstract**: Data donation, an emerging user-centric data collection method for public sector research, faces a gap between participant willingness and actual donation. This suggests a design absence in practice: while promoted as "donor-centered" with technical and regulational advances, a design perspective on how data choices are presented and intervene on individual behaviors remain underexplored. In this paper, we focus on pre-donation data exploration, a key stage for adequate and meaningful informed participation. Through a real-world data donation study (N=24), we evaluated three data exploration interventions (self-focused, social comparison, collective-only). Findings show choice framing impacts donation participation. The "social comparison" design (87.5%) outperformed the "self-focused view" (62.5%) while a "collective-only" frame (37.5%) backfired, causing "perspective confusion" and privacy concerns. This study demonstrates how strategic data framing addresses data donation as a behavioral challenge, revealing design's critical yet underexplored role in data donation for participatory public sector innovation.

## 1. Introduction

Data donation, the voluntary transfer of personal data for research or public benefit, has emerged as a promising method for participatory public sector innovation. By granting individuals control over what they share, data donation enables access to granular, private, and retrospective personal data while respecting user autonomy (Ohme et al., 2024). However, despite being celebrated as user-centric, data donation faces a critical challenge: significant gaps between donation willingness and actual behavior (Gómez Ortega et al., 2021; Hase et al., 2024). Current implementations remain disconnected from donor experiences, focusing primarily on technical systems and policy guidelines for researchers rather than producing better experiences for donors (Groot Kormelink et al., 2025).

This challenge is central to the pre-donation data exploration stage, where participants review their personal data before deciding whether to donate. While multiple visual tools are implemented to support comprehension and adequately informed consent (Boeschoten et al., 2023; Franzen et al., 2024; Pfiffner et al., 2024), they often use standardized, impersonal designs with generic interface elements regardless of data type, leaving participants struggling to interpret implications (Groot Kormelink et al., 2025). We argue this stems from treating data exploration as a technical usability problem rather than a behavioral design problem. Research on traditional donations has demonstrated that framing—such as social comparison information (M.DiCosola & Neff, 2022) or appeals to collective benefit (Hoover et al., 2018)—powerfully moderates behavior. Yet data donation has largely overlooked these dimensions, defaulting to showing participants only their own data in isolation.

We investigate how different presentations of personal data during exploration influence donor behavior and decision-making. Through a controlled study (N=24) examining calendar data donation, we designed three exploration interventions: an Individual-Only Exploration (baseline), an Individual-Plus-Collective Exploration, and a Collective-Only Exploration. Our findings reveal pre-donation exploration is not neutral but a critical intervention: the same data, framed differently, improved donation rates from 37.5% to 87.5%—a 50-percentage-point gap.

Our study makes three contributions. First, we provide empirical evidence that pre-donation exploration is a high-impact behavioral intervention. Second, we contribute an empirically validated social framing design pattern, demonstrating that grounding participants in personal data before introducing peer comparison significantly improves donation outcomes. Third, we propose data donation as a systematic behavioral design problem, arguing that effective choice design must be orchestrated across interface presentation, contextual framing, and policy constraints.

## 2. Background and Related Work

### *2.1 Data Donation as an Emerging Public Sector Method*

Public sector practitioners increasingly use design to shape policies and services, moving beyond bureaucratic traditions toward human-centered administration (Bason & Austin, 2022; Mortati et al., 2022). As governments face “wicked” and complex challenges, there is a growing need for participatory innovations to connect citizens with society and institutions into closer dialogue. Digital transformation particularly demands high-quality, ethically sourced civic data to inform policy and reimagine public services, creating a clear need for new, citizen-centric data collection methods.

Data donation is an emerging process where individuals voluntarily transfer their personal data to an entity, such as a research organization, for a specific, informed purpose (Carrière et al., 2024). This method has grown due to the ubiquity of digital services generating vast personal data, and significantly enabled by the 2018 GDPR right to data portability (Art.20 GDPR, 2025; Gómez Ortega et al., 2023)

As a data collection method, data donation is valued for its user-centric characteristics (Ohme & Araujo, 2022). Unlike aggregate data collection approaches such as web scraping or API access, data donation grants individuals control over what they share. This yields key advantages: donated data is highly granular at an individual level, can include private data inaccessible through platform APIs, and enables retrospective collection (Breuer et al., 2021; Ohme et al., 2024). This potential has attracted interdisciplinary interest spanning technical domains such as privacy-preserving data management (Boeschoten et al., 2022), psychological research on donation motivations and willingness (Keusch et al., 2024; Skatova & Goulding, 2019), policy-oriented practice guideline (Carrière et al., 2024), organizational perspectives on infrastructure (Liu et al., 2025; Qiu et al., 2026) and legal implications under privacy regulations (Breuer et al., 2021; Janssen et al., 2025; Sloan et al., 2020).

While celebrated as user-centric, data donation processes often remain disconnected from donor experiences (Groot Kormelink et al., 2025). Most current implementations focus on advancing technical systems and policy guidelines for researchers rather than producing better experiences for donors, leading to high dropout rates and significant gaps between willingness to donate and actual behavior (Gómez Ortega et al., 2021; Hase et al., 2024). Values such as trust, solidarity, and public good value, which are aspects of collective commitment contingent on recognizing shared stakes with others (Prainsack & Buyx, 2017), have been found as key motivational preconditions for data sharing (Skatova & Goulding, 2019). Design research offers a productive lens for addressing this gap: recent participatory design scholarship has foregrounded how meaningful citizen engagement with digital public services requires attending to how people experience participation, not just whether participation occurs (Christiansson et al., 2024; Kirk et al., 2024; Chen & Schmidt, 2024; Chen et al., 2025). Applied to data donation, these call for approaches that treat the donation journey itself as a site of participatory intervention, where donors are active agents rather than passive subjects (Gómez Ortega et al., 2024).

## *2.2 Pre-Donation Data Exploration Challenge*

A core challenge in data donation is achieving "adequately" informed consent that enables participants to make truly informed decisions about their personal data sharing (Gómez Ortega et al., 2023). Participants typically need to prepare and transmit their personal data files in machine-readable formats such as JSON or CSV, which may contain more data than the data collector requires or highly sensitive personal information (Boeschoten et al., 2022). The technical nature of donating data, and the sheer volume of many datasets that are valuable in donation contexts, make data donation processes difficult for participants to comprehend (Ohme et al., 2024), complicating meaningful informed consent (Groot Kormelink et al., 2025).

To address this, researchers have increasingly emphasized data exploration as a critical pre-donation stage. More robust workflows have emerged involving local device processing (Boeschoten et al., 2022), inspection and deletion options (Boeschoten et al., 2023; Pfiffner et al., 2024), and interactive interpretation assistance (Franzen et al., 2024; Gómez Ortega et al., 2024). This stage increasingly adopts data visualizations, such as line graphs, bar charts, and word clouds, to make the contents of data files legible and comprehensible (Boeschoten et al., 2023). Meanwhile, increasing research finds that visual previews could potentially reduce privacy concerns and increase transparency (Alhadad, 2018; Ohme & Araujo, 2022).

However, current data exploration often adopts a one-size-fits-all design that fails to account for diverse contexts and donor needs. Most implementations use generic interface elements and visualization templates, regardless of whether donations concern social media behavior, health information, or financial data (Boeschoten et al., 2023; Groot Kormelink et al., 2025; Keusch et al., 2025), ignoring individual differences in data literacy, privacy preferences, and personal contexts (Gómez Ortega et al., 2021). Simply adding visualizations does not address comprehension issues; participants still struggle to interpret data and donation implications (Groot Kormelink et al., 2025).

This narrows the broad donor-centricity problem to a specific gap at the pre-donation stage: it remains unclear what alternative data exploration designs might work, and little is known about participants' experiences of the data donation process (Groot Kormelink et al., 2025). Current uniform, template-based data exploration approaches suggest that this crucial stage has not been designed with sufficient attention to donor perspectives, needs, and decision-making.

## *2.3 Self-Focused Default and Underexplored Alternatives*

The design norm for pre-donation data exploration shows participants their own data in isolation—a self-focused exploration design consistently adopted across major data donation applications and services, such as Port (Boeschoten et al., 2023), Data Donation Module (Pfiffner et al., 2024), and the Tidepool Data Platform.  This default stems from literal interpretation of informed consent principles and aligns with impersonal analytical traditions rather than human-centered relational approaches (Bason & Austin, 2022). However, this design norm may be problematic for several reasons grounded in data visualization research and behavioral science.

First, data visualization functions as a powerful rhetorical act that significantly affects interpretation and decision-making (Hullman & Diakopoulos, 2011; Markant et al., 2023), creating an illusion of objectivity (Kennedy et al., 2016). This is especially concerning in “high-stakes” consent processes, which govern medical, financial, bodily, and other intimate data donation activities. Research in data donation context further finds participants interpret data visualizations as objective representations even when data was incomplete or contradict their experiences, narrativizing rather than questioning counterintuitive data (Groot Kormelink et al., 2025).

Second, behavioral science research on traditional donations (charitable, blood, organ, etc.) reveals that framing and choice presentation powerfully moderate donation behavior (Hoover et al., 2018; Keser et al., 2023). Social and in-/out-group norms are a particularly strong force in encouraging “appropriate” behaviors (Anderson & Dunning, 2014; Fehr et al., 2002). Factors such as social comparison information, beneficiary framing, and collective versus individual perspective appeals significantly influence participation behaviors and donation decisions (André et al., 2017; Feine et al., 2023; Sneddon et al., 2020), suggesting similar framing effects may operate in data donation.

Yet despite these insights from traditional donation research, data donation research lacks systematic investigation of choice design and framing effects. Existing findings are scattered and contradictory: while Skatova and Goulding (2019) find collective benefit appeals positively predict donation intention and personal benefit appeals negatively predict it, studies of practical donations find personal benefits like gaining self-insights are key motivators (Gómez Ortega et al., 2023; Sörries et al., 2024). While behavioral science explains why framing effects operate, design research provides the methodological basis, like Choice Triad framework (Schmidt et al., 2022) and systemic participatory approach (Blomkamp, 2022), that intervene through exploratory and participatory processes to situate citizen experience at the centre of public sector innovation. This study sits at the intersection of behavioral and public participatory design, examining (1) how pre-donation data exploration designs influence donors' donation behavior, judgment, and decision-making, while exploring (2) the design possibilities of choice framing for data donation participation.

# 3. Method

To answer this research question, we conducted a user study comparing three data exploration interventions in a real data donation practice. Following a Research through Design (RtD) approach (Zimmerman et al., 2007), we designed and developed a data exploration-donation platform as the primary vehicle for knowledge generation; guided by the established data donation practice guideline (Carrière et al., 2024), we emphasized meaningful donation purpose and adequate informed consent.

The study was approved by Illinois Tech’s Institutional Review Board (#IRB-2025-47). All participants provided full informed consent to participate in the data exploration study and fully understood that consent to participate in the study was distinct from consent to donate their data. Regardless of participants' post-exploration decisions to donate or decline, no personally identifiable data was stored after study sessions.

## *3.1 Scenario: Institutional Calendar Data Donation*

We conducted the study in the context of an institutional calendar data donation (Google Calendar) at the Institute of Design (ID) at Illinois Tech in Chicago, USA. Calendar data represents a valuable resource for academic institutions seeking to understand scheduling patterns, facilities use, resource allocation, and community engagement, while also containing personally sensitive information. For students, calendars include academic commitments (classes, advisor meetings), collaborative activities (group projects, student organizations), and personal appointments—making this data type both institutionally valuable and personally sensitive. Participants were presented with an opportunity to help the institute better understand scheduling patterns, optimize facility usage, and improve academic planning by donating individual data. While more bounded than broader public good causes common in health or civic data donation, improving institutional scheduling represents a shared community value where trust and solidarity motivations remain relevant.

## *3.2 Interventions: Baseline and Alternative Data Exploration Design*

We designed three interventions that varied data exploration presentation from individual-focused to collective-focused, based on Skatova and Goulding’s identification of personal and collective benefits as motivations for data donation (Skatova & Goulding, 2019). Starting from a default, self-focused, and individual-only exploration, we developed two alternatives to explore how different presentations of personal data would affect participants' experience and subsequent decisions.

- **Group A** (*Baseline: Individual-Only Exploration*): This intervention followed the "default design" norm, focusing on individual self-reflection by presenting participants' personal data compared only to their own patterns.
- **Group B** (*Alternative 1: Individual-Collective Exploration*): This intervention introduced social comparison by contrasting participants' individual data against an anonymized, aggregated pool of data from other donors.
- **Group C** (*Alternative 2: Collective-Only Exploration)*: This intervention emphasized collective contribution by integrating participants' personal data into the collective pool and displaying only anonymized, aggregated information.

While all three interventions meet the same informed consent requirements, they reflect different degrees of social framing representing a spectrum from individual to institutional framing: Group A foregrounds purely individual decision-making, Group B introduces an institutional-social dimension through peer comparison, and Group C situates the decision entirely within a collective institutional context. The core difference between interventions was the informational context provided. Table 1 illustrates how the same data point would be annotated across the three groups.

*Table 1 Overview of three data exploration intervention examples.*

| Grp | Intervention Design | Data Comparison | Example Annotations | Example Visualizations |
|---|---|---|---|---|
| A | **Individual-Only Exploration** | Personal data (mine) *vs.*<br>Personal data (mine) | "The Fall 2024 semester has 112 days. YOU had events on *[n]* days. YOUR longest scheduled day was *[date]*..." | |
| B | **Individual-Collective Exploration** | Personal data (mine) *vs.*<br>Collective data (theirs) | "The Fall 2024 semester has 112 days. YOU had events on *[n]* days; OTHER STUDENTS averaged *[n]* days. YOUR longest scheduled day was *[date]*; for OTHER STUDENTS..." | |
| C | **Collective-Only Exploration** | Collective data (ours) *vs.*<br>Collective data (ours) | "The Fall 2024 semester has 112 days. STUDENTS had events on *[n]* days. The longest scheduled day for STUDENTS was *[date]*..." | |

## *3.3 Exploration Platform and Visualization Archetypes*

To ensure ethical and adequate informed consent, we followed established informing templates (Gómez Ortega et al., 2023; Sloan et al., 2020) that treat consent as a process, requiring data exploration to inform participants about how data is collected, how they can access their data, what exactly is in their data, what makes data personal to them, and what data they are donating.

To fulfill these requirements while enabling controlled comparison across interventions, we designed and developed a privacy-preserving web platform implementing four complementary visualization archetypes. Each archetype addresses specific informational needs, and together they provide complete exploration coverage. Participants signed in with their institutional Google Calendar credentials to extract their data, then scrolled through the four visualization archetypes sequentially. Raw data in both table and JSON formats were also available for optional exploration (see Figure 1).

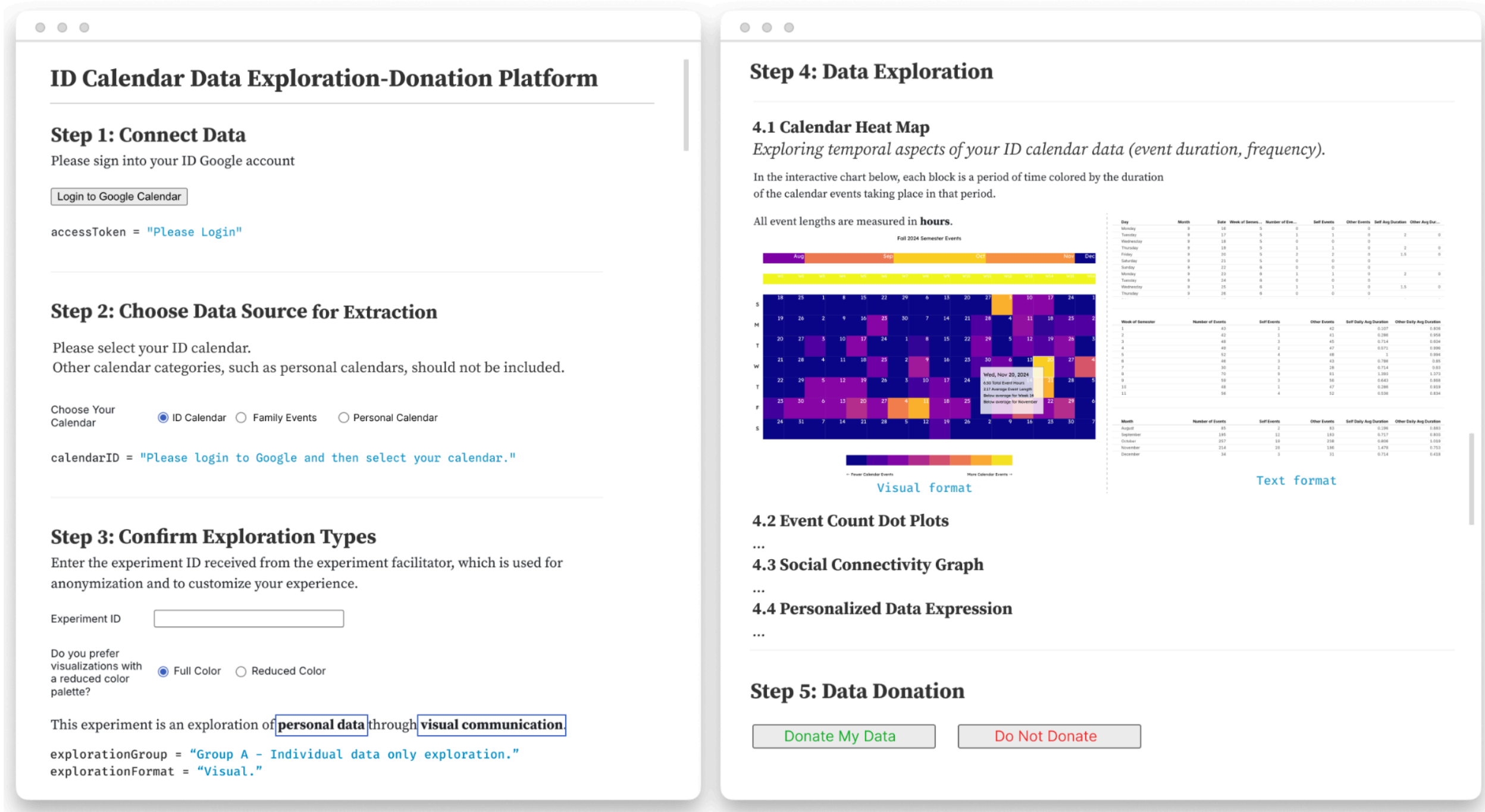

*Figure 1* *The data exploration and donation platform developed for the ID calendar data donation project, showing Steps 1–3 (left) and Steps 4–5 covering data exploration and donation decision-making (right). In Step 3, participants were informed of their assigned exploration group (A/B/C) and presentation format (visual/text) through both platform messages and verbal explanation from the researcher. In Step 4, participants were required to view all four visualization archetypes before proceeding to Step 5. After Step 5, regardless of their donation decision, all participants were guided to log out and verify that the platform retained no data. Participants could enable multiple accessibility accommodations at any point throughout the data exploration, including color vision, contrast, and text size options.*

The four archetypes were designed to collectively ensure adequate informed consent coverage, following data visualization design principles of perceptual clarity, layered annotation, and progressive disclosure to support comprehension across varying levels of data literacy (Franconeri et al., 2021). Each archetype addresses a distinct dimension of the calendar data through varied visual representations, annotations, and levels of interactivity (Hullman & Diakopoulos, 2011). Critically, visual representations and interactive affordances were held constant across all three intervention groups; only the data comparison and annotations (see Table 1) within each archetype varied, isolating informational framing as the independent variable while controlling for visualization format effects (see Figure 2):

(1) *Calendar Heat Map:* A temporal visualization showing event frequency and duration across a previous academic semester, allowing participants to explore patterns in their schedule.

(2) *Event Count Dot Plots*: Category-based visualizations displaying distributions of meeting types, locations (online/in-person), and other calendar event attributes.

(3) *Social Connectivity Graph:* A network visualization revealing connections between calendar events and attendees, highlighting social dimensions of calendar data.

(4) *Personalized Data Expression*: An interactive visualization allowing participants to select variables of interest and create custom visual metaphors to represent their calendar data aesthetically.

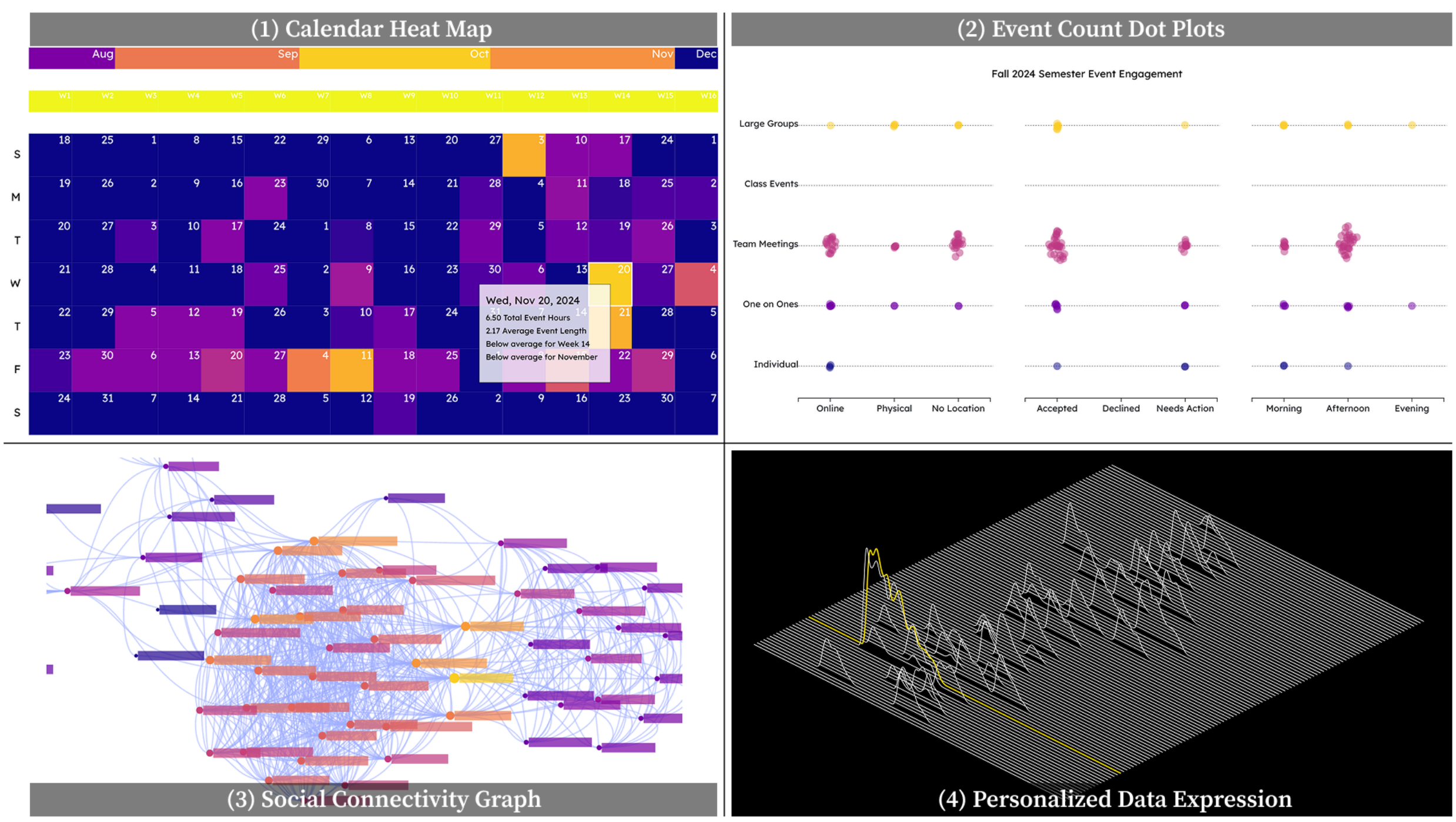

*Figure 2 Examples of the four visualization archetypes from the baseline exploration design (Group A: individual-only exploration)*

## *3.4 Procedure*

The study procedure followed Carrière et al.'s guidelines (2024) and a structured "consent-as-a-process" procedure example of in-person session (Groot Kormelink et al., 2025)(Figure 3):

1. **Consent to Exploration:** Participants consented to participate in the data exploration study and allow temporary, local processing of their calendar data. They were explicitly informed no data would be collected by default. Introduction materials (IRB overview, donation context, platform information) were presented through printed documents and researcher's oral explanation.

2. **Pre-Study Survey:** Participants completed a survey evaluating their prior understanding of both data and donation projects based on traditional informed consent materials (from Step 1), before any data exploration.
3. **Data Extraction:** Participants logged into our platform with their institutional Google Calendar credentials. Data was processed locally in their browser to generate their assigned intervention (A, B, or C).
4. **Data Exploration:** Participants explored their data through four interactive visualization archetypes while providing think-aloud commentary. Researchers could not view participants' screens to ensure privacy.
5. **Donation Decision:** Participants scrolled to the bottom of the platform after exploration to make their donation choice:"Donate My Data" or "Do Not Donate."
6. **Verification:** Participants logged out and verified that the platform retained no data. Those who donated verified their donated data package was properly anonymized.
7. **Post-Study Survey & Follow-up Interview:** Participants completed a post-study survey (mirroring pre-study questions) and a semi-structured interview discussing their experience.

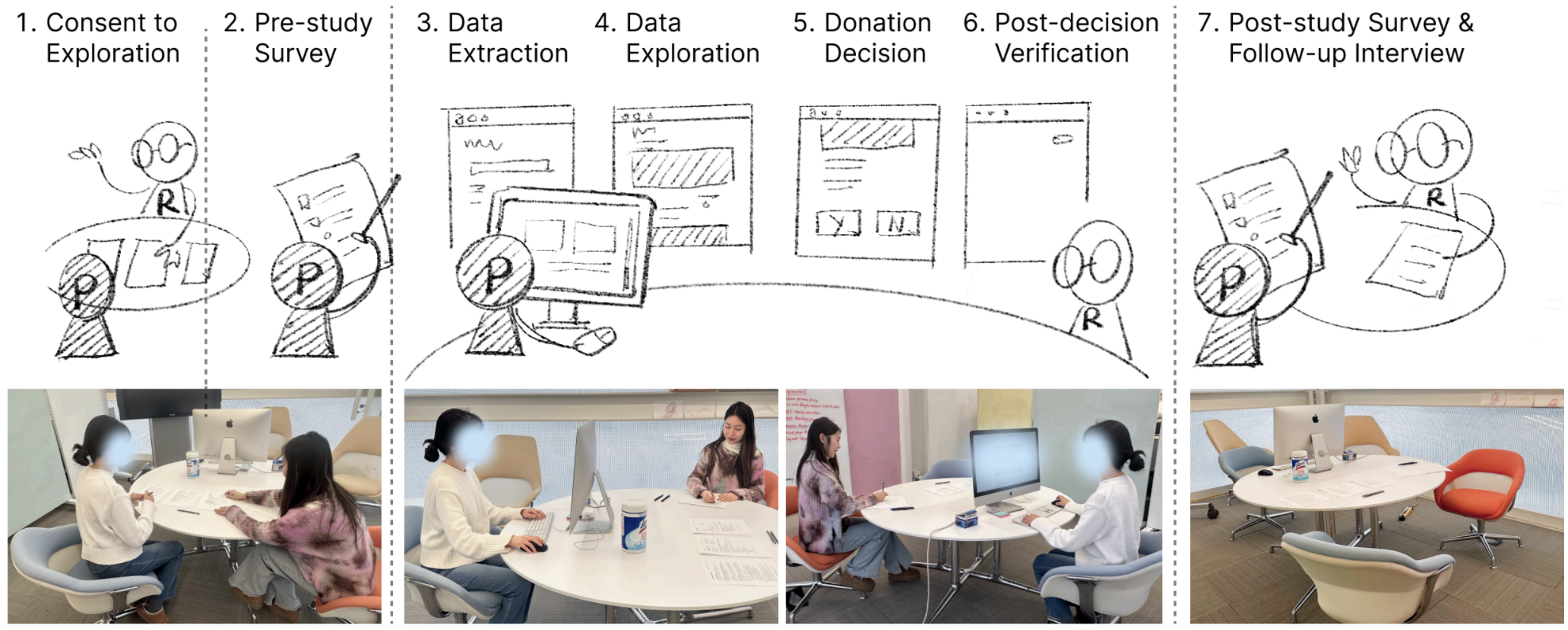

*Figure 3 Overview of in-person data donation procedure and environmental setting of the user study.*

## *3.5 Participants*

We recruited 30 student participants from the Institute of Design (ID) of approximately 80–100 individuals (30-38% representation rate): 24 for the main study and 6 for pilot testing. This sample represents nearly a third of the defined target population, making it a meaningful representation for a bounded institutional study. All participants volunteered to participate in the study without compensation.

These 24 participants were equally divided into the three intervention groups (n=8 per group). Within each group, 6 participants experienced the visual archetypes while 2 received a text-based control showing identical information in table format. Given prior research demonstrating visualization's equivalent efficacy for data exploration (Keusch et al., 2025), we focused on comparing informational design across interventions rather than reaffirming visualization benefits.

### 3.6 Data Collection and Analysis

We collected both quantitative data (donation decisions, exploration duration, pre/post-survey ratings) and qualitative data (think-aloud protocols, open-ended survey responses, semi-structured interviews). Analysis followed a mixed-methods approach: descriptive statistics for quantitative data and thematic analysis (Braun & Clarke, 2019) for qualitative data. We analyzed patterns within and across groups to understand how each intervention influenced participants' behavior (donation outcomes and engagement), judgment (understanding and helpfulness perceptions), and decision-making (donation rationale).

## 4. Results and Findings

### 4.1 Behavior: Donation Results

Group B (individual-collective exploration) showed the highest donation rate at 87.5% (7/8 participants), followed by Group A (individual-only exploration) at 62.5% (5/8), and Group C (collective-only exploration) at 37.5% (3/8), seen in Figure 4. This 50-percentage-point difference between Group B and C represents a substantial practical difference with medium-to-large effect size (Cramer's V = 0.42). Comparison of visualization (n=18) and text-table (n=6) conditions revealed no substantial differences in donation decisions ($R^2$ = 0.934), indicating presentation format explained minimal variance in outcomes. Qualitative data showed no strong preference for specific visualization archetypes, as Participant a3 noted, "I do not think any visualization was particularly impactful as much as the exercise as a whole" (a3). These findings informed our decision to merge participants within their respective intervention groups (A, B, or C) and focus analysis on informational design — how the data was contextualized and presented.

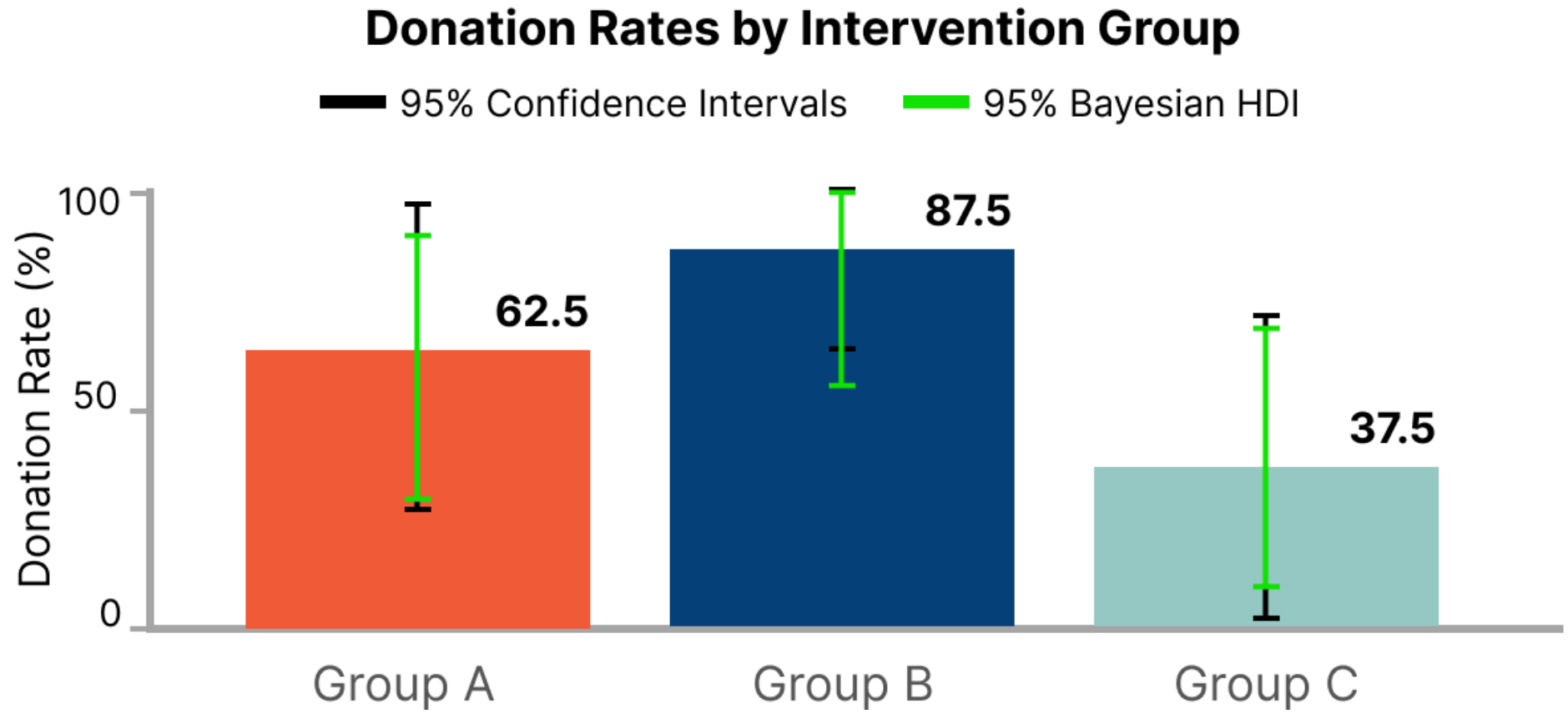

*Figure 4 Bar chart of donation rates by intervention group. Bayesian analysis confirms these differences are robust despite the small sample: the probability that Group B outperformed Group C was >99.9%, supported by both 95% confidence intervals (black) and 95% highest density intervals (green).*

## *4.2 Decision-Making: Rationale for Donating*

We investigated donation reasoning through audio transcripts and post-study surveys, conducting thematic analysis to identify how participants in different groups rationalized their decisions.

Groups A and B (62.5% and 87.5% donation rates respectively) exhibited markedly different reasoning patterns than Group C (37.5%), despite all groups receiving identical information about data use and privacy. Participants in Groups A and B more frequently emphasized enabling factors such as trust in researchers ("I trust the Institute of Design and its researchers" (a6) and "I trust that my data is in good hands" (b7)), low data sensitivity ("My institutional calendar doesn't really contain personal events" (a8)), and perceived transparency. They also valued the exploration process, with visualizations helping them "understand what data I was donating" (a1). Notably, Group B participants demonstrated the strongest community-oriented reasoning, expressing desires to "help make better digital experience for the community" (b1) or enable "better schedule decisions" (b2). This may reflect their exposure to both personal and peer data contexts, which appeared to strengthen their sense of collective contribution.

In stark contrast, Group C participants, where the majority did not donate, predominantly emphasized concerns and barriers. Privacy concerns emerged as their primary preoccupation, with participants worried data might be "too personal" (c8) or expressing uncertainty about the "safety of data donation in general" (c4). The collective data view appeared to make potential risks more salient than personal connections to the outcomes of the research. Group C participants also expressed greater uncertainty about what would be collected, who might access it, and how donation benefits others. Unlike Groups A and B, where privacy was mentioned hypothetically, Group C participants framed privacy as an active barrier to donation. Even participants who trusted the immediate researchers distinguished this from concerns about broader data access beyond the research team.

Privacy concerns were mentioned across all groups, suggesting this remains a fundamental consideration in data donation decisions regardless of framing. However, the balance between enabling factors and barriers differed substantially: Groups A and B reasoning leaned toward trust, transparency, and community benefit, while Group C reasoning centered on privacy concerns and uncertainty. These patterns suggest that how personal data is framed during exploration significantly shapes not only the decision to donate but also the underlying reasoning participants use to justify those decisions.

## *4.3 Judgment: Intended vs Adopted Perspective During Donating*

To investigate participants' actual adopted exploration perspectives, the authors inductively derived four perspective codes from think-aloud protocols and post-task interviews through thematic analysis. Codes were developed by grouping recurring interpretive patterns based on how participants narrated whose data they perceived themselves as viewing — their own data, others' data, or the data about the collective — and whether they expressed uncertainty about this. Table 2 presents the resulting codes applied across groups.

*Table 2 Adopted perspective codes by group*

| Code | Definition | Example | A | B | C | Total |
|---|---|---|---|---|---|---|
| **Self-focused perspective** | Primarily interpreting data in terms of oneself | *"I didn't realize how much time I spent on this project meeting."* | **10** | 3 | 2 | **15** |
| **Social comparative perspective** | Comparing one's data to others or to averages | *"So.. I'm above average compared to my classmates."* | 1 | 3 | 1 | 5 |
| **Collective-only perspective** | Focusing on community or collective patterns | *"It looks like most students use it in the morning"* | 0 | 2 | 3 | 5 |
| **Confused perspective** | Mixing perspectives or unsure whose data is whose | *"Wait, is this mine or everyone's?"* | 0 | 0 | **6** | 6 |

The predominance of self-focused interpretation across groups (15 of 31 codes) reveals an intuitive data reading pattern: people naturally begin by focusing on what is most salient to them by locating themselves in the data. Group A participants overwhelmingly interpreted visualizations as personal reflections, while Group B participants typically started self-focused but shifted toward collective orientation. For example, Participant b10 first reported their personal learning, "I didn't realize how my individual calendar can link to other people!"; later expanded their reflection: "... so Google Calendar is not just individual, the features can become highly social and powerful for the Institute of Design..." (b10). This suggests the creation of dual personal and peer relational contexts, enabling perspective shifts from initial self-focused insights to broader communal reflections.

However, we observed that this intuitive drive to locate oneself becomes problematic when only collective data is presented. Despite being explicitly informed of their assigned perspective through both textual labels on the platform interface and oral reminders from researchers, Group C participants were dominated by perspective confusion. Many struggled to interpret whose data they were viewing, with Participant c4 persistently asking why they were "not seeing what I potentially might contribute" (c4). This suggests that confusion amplifies perceived risks and erodes trust, revealing a fundamental tension: genuine collective perspectives require complete anonymization—no individual can be identified, not even oneself—yet data donation inherently requires understanding one's own personal contribution.

## 4.4 Experience with Data Exploration

From post-study surveys and interviews, pre-donation exploration was universally perceived as helpful for decision-making, though helpfulness alone did not predict data donation decisions. However, the groups' differing understanding shaped divergent emotional responses to the exploration. While Groups A and B expressed "surprise" and "fascination" (a8, b2, b7, b8) and "delight" (a5, a6, b4) at visualizing their data, Group C participants struggling with interpretability reported "alarm," "discomfort," and "fear" ("I might stop putting my personal calendar data on Google Calendar now!" (c1) and anxiety that they had "never thought about" data access risks before (a7)). This pattern reveals that in sensitive data contexts, confusion erodes trust and amplifies privacy concerns. As Participant a1 noted, initial trust from transparency was quickly eroded by emotional reactions and cognitive load, such as privacy fears and exhaustion from reading (a1).

# 5. Discussion

The results of our study reveal that the design of the pre-donation data exploration is not a neutral feature but a critical, high-impact intervention. Our findings present a clear puzzle: while the modality of presentation (visualization vs. text) and the specific visualization archetype (heatmap, dot plot, network diagram) had no discernible impact found on participant decisions, the informational design of how data was contextualized dramatically altered behavior. This suggests that design for data donation is not a simple UI/UX or usability problem; instead, it is a systematic behavioral design problem centered on framing data choices for participatory public sector innovation.

## *5.1 Behavioral Effects Beneath "Adequately" Informing*

To date, the practical discourse on data donation has largely centered on technical advancement and ethico-legal frameworks (Groot Kormelink et al., 2025). The primary focus has been on achieving "adequately and meaningful informed consent" through maximum transparency (Gómez Ortega et al., 2023), typically implemented by objectively exposing all data and its implications in some standard tables or figures to inform participants (Boeschoten et al., 2023). This approach, which we implemented as our Group A (baseline), implicitly frames the donor as a rational actor, assuming that if participants are simply presented with all information, they will make a logical decision.

However, our study proves this standardized adequate informing approach is only partially effective. In our study, we strictly adhered to these established norms for informing and consenting participants across all three groups. The significant divergence between the highest (Group B: 87.5%) and lowest (Group C: 37.5%) data donation rates appeared to be driven by how the data was contextualized and framed across three social comparison conditions, and suggests the value of further exploration into several behavioral dimensions of data donation:

- **The Collective Backfire** (Group C), which framed the data as an abstract collective, caused a significant drop to a 37.5% donation rate. The design, intended to be pro-social, instead caused reluctance, with participants feeling "lost" (c6) and "scared" (c6, c1). This confusion directly amplified privacy concerns, suppressing the very altruism it was meant to inspire.

- **Social Comparison as Leverage Point** (Group B), in contrast, achieved an 87.5% donation rate, demonstrating a well-designed behavioral intervention. By grounding the participant in their own data first (personal relevance) before contextualizing it against peers (pro-social context), it fostered a sense of shared, communal action ("...it spoke to me how connected we are as students to the school as a whole. It is powerful to know how connected to others we can be" (b8)). By social comparison,

> this group expressed the strongest altruistic motives and highest understanding of the donation context that the "adequately and objective" transparency (Group A) unaddressed. This mirrors Prainsack and Buyx's (2017) account of solidarity as requiring recognition of shared similarity before collective commitment becomes possible — a precondition Group C's abstract collective framing bypassed.

Behavioral impacts of framing are not optional; they are unavoidably embedded in any choice presentation. But data donation contexts also contain a wide array of other behavioral issues — gain-loss comparisons, risk interpretation, immediate versus delayed value — that suggest the potential value of casting a wider behavioral net in the interest of designing donation systems that are not only standardized and effective, but also contextualized and ethical.

## *5.2 Framing Through Data Contextualization*

Behavioral science typically conceptualizes framing as a message-level phenomenon. How a choice is presented through wording, emotional tone, or positive versus negative narratives can significantly alter decisions, often more than the choice itself (Tversky & Kahneman, 1981). In traditional donation contexts, such as charity, blood, and organ donations, framing studies largely focus on appeal framing, examining how messages or evocative text (sometimes paired with static images) influence empathy, urgency, or perceived social obligation (Fagley et al., 2010; Feine et al., 2023; Hoover et al., 2018).

Our study extends framing theory by demonstrating that in data-rich contexts, framing operates through data contextualization itself, not just textual appeals. The social framing intervention—positioning personal data through Self-Focused Framing, Social Comparison Framing, or Collective-Only Framing—functioned as the primary behavioral lever. This establishes social comparison for data contextualization as a validated, replicable design pattern for data donation systems. Social Comparison Framing, which grounded participants in personal data before introducing peer comparison, proved most effective.

While our study validates social framing through data exploration as a high-impact intervention, this represents only one dimension of how framing may operate in data donation contexts. Future research could investigate whether framing operates through additional modalities beyond data contextualization, such as visual emphasis, interaction affordances, or temporal structure. These insights extend beyond pre-donation exploration to recruitment messages, consent screens, and procedure flows throughout the donation journey, offering designers opportunities to stimulate meaningful participation while also assigning them a new responsibility to ensure that the framing design does not compromise informed consent or exploit behavioral tendencies.

## *5.3 Designing Data Choices for Public Participation*

Recent scholarship in public sector design has emphasized the need to move beyond interface-level interventions toward systematic approaches that address complex behavioral challenges (Bason & Austin, 2022; Junginger, 2017). Schmidt et al. (2022) propose a systematic behavioral framework distinguishing three design levels: choice posture (individual preferences), choice architecture (contextual framing), and choice infrastructure (policy constraints). Though originally developed for public health policy, this framework applies directly to data donation practices: how donation choices are designed shapes who participates, and how well data donation can function as a mechanism for participatory public sector innovation. Our study demonstrates why this multi-level approach is essential. We found that transparency acts as a double-edged sword. While exploration surfaced what data was collected and how it would be processed, it simultaneously created new anxieties, with participants noting they had "never thought about who already has access to my data" (b8) before.

Group A (Baseline) exemplified choice posture design (Schmidt et al., 2022), presenting objectively transparent choices at the UI level in an impersonal manner. While achieving adequate results (62.5% donation rate), this design failed to address transparency-induced anxieties, relying entirely on pre-existing trust. This reveals a fundamental limitation of choice posture-only approaches: while individual factors such as trust or data literacy inevitably shape donation decisions, designing purely around individual preferences is neither scalable nor reliably effective. Some participants donate because of trust, yet recent scholars also suggest that non-trust may produce more sustainably informed participation in the long term (Ponti et al., 2024). Rather than endlessly customizing interfaces to individual postures, data donation design needs to move upward, supporting individual choices through deliberate choice architecture that can work across diverse participants.

Group B's highest donation rate demonstrates the power of this shift. By embedding social norms into transparency frameworks, Group B communicated community-situated context rather than just presenting data. Critically, it was the only condition showing no negative changes across any understanding dimension, improving comprehension while mitigating anxieties. As Participant b3 noted: "it spoke to me how connected we are as students to the school as a whole" (b3). Social comparison as a strategy for civic participation is well-established: Allcott's (2011) evaluation of the Opower program showed that comparing household energy use to similar neighbors reduced consumption by 1.4–3.3% across 17 experiments. The mechanism mirrors Group B conditions, which grounded individuals in their own data first, then contextualized it through peer comparison, activating shared norms and community contribution in ways that abstract collective appeals cannot. Our study indicates that social comparison can also be operationalized through data visualization itself — a novel modality for choice architecture in data-rich contexts. Since visualizations are never neutral but always rhetorical acts that shape interpretation (Hullman & Diakopoulos, 2011; Kennedy et al., 2016), this suggests that proactively embedding behavioral strategies into visualization design, rather than passively presenting data under an assumption of objectivity, might further improve the choice architecture of data donation.

However, Group C's lowest donation rate (37.5%), combined with reported confusion and the weakest understanding improvements, suggests that regulatory requirements for data donation suffer from choice infrastructure challenges. For example, privacy regulations required complete anonymization, making data unintuitive to interpret despite the fact that Group C's visualizations most accurately represented how researchers would work with data. This points to the need for choice infrastructure design (Schmidt et al., 2022): designing within policy constraints while maintaining comprehensibility. GDPR's right to data portability (Article 20) shows how infrastructure can enable citizen-driven data participation by establishing legal and technical standards (Art. 20 GDPR, 2025). Yet our findings expose a gap this infrastructure did not anticipate: the same anonymization requirements designed to protect donors can undermine the personal recognition that makes meaningful participation possible. Just as dark pattern regulations established frameworks to govern manipulative interface design (Chen et al., 2025; Luguri & Strahilevitz, 2021), data presentation in donation contexts warrants analogous regulatory attention. For instance, individual exploration stages must clearly surface to users where and how their own data appears within the collective, addressing both data privacy requirements and donors' psychological need for personal recognition amidst uncertainty.

Taken together, our findings argue that choice presentation in data donation is not a singular goal but a multifaceted challenge requiring framing design across all three levels. Simple disclosure is insufficient; successful data donation design must anticipate the anxieties that transparency itself creates while building genuine understanding within regulatory constraints.

## 6. Limitations and Future Work

Our findings are exploratory, limited by a small (N=24), single-institution sample, and a low-sensitivity data type (Google Calendar). The "social comparison" frame's success (Group B) may not generalize and requires re-testing with larger, diverse populations and high-stakes data, where these framing approaches could plausibly backfire. Future work should also test other behavioral levers (such as gain-loss frames, immediate-delayed values) and address the key tension, revealed by Group C, between privacy accommodations (like anonymization) and user comprehension. Finally, our study focuses specifically on informational framing effects and does not account for broader motivational and social factors, such as trust in institutions, solidarity motivations, public data literacy, peer norms, and cultural attitudes toward data sharing. This includes how critical awareness or non-trust toward data processes might support more informed long-term engagement, both of which warrant investigation in future work. Future research should also measure data literacy pre- and post-exploration, both as a control variable and to examine whether the exploration process itself influences participants' data literacy.

## 7. Conclusion

This study makes three primary contributions to the fields of public sector design and data donation: First, we provide empirical evidence that pre-donation data exploration is a critical, high-impact intervention. Behavioral effects are already embedded in commonly used "default" designs (Group A), while alternative designs meeting the same informing requirements can produce dramatically different outcomes, ranging from 37.5% to 87.5% donation rates.

Second, we contribute an empirically-validated "social framing" design pattern and explore framing's multimodal potentials towards data presentation. The Individual-Collective comparison frame (Group B) succeeded by grounding participants in personal data before introducing peer comparison, while the Collective-only frame (Group C) backfired through abstract anonymization that caused confusion. Beyond social framing, we identified multiple potential framing modalities including visual emphasis, interaction affordances, and temporal structure, extending beyond traditional textual message framings.

Third, we propose data donation as a systematic behavioral design problem for public sector innovation. Our findings argue that orchestrated choice design across interface presentation, contextual framing, and policy alignment, opening new directions for creating participatory data infrastructures that are both compliant and empowering for citizens.

**Acknowledgements:** The authors are grateful to all participants from the Institute of Design at Illinois Institute of Technology whose willingness to share their time and personal data made this research possible.

## About the Authors:

**Zeya Chen** is a PhD candidate at the Institute of Design at Illinois Tech. Her research focuses on systemic behavioral design, personal data participation, and human-AI interaction. Her work spans UX research, AI product management, and privacy-centered design tools.

**Zach Pino** is an Assistant Professor at the Institute of Design at Illinois Tech. His research focuses on inclusive and creative uses of data in the design process. His work spans data visualizations, generative algorithms, data capture instruments, and accessible interactive experiences.

**Ruth Schmidt** is an Associate Professor at the Institute of Design at Illinois Tech, whose research and project work sits at the intersection of behavioral science, humanity-centered design, and complex systems.